\documentclass{article}%
\usepackage{amsmath}
\usepackage{amsfonts}
\usepackage{amssymb}
\usepackage{graphicx}%
\providecommand{\U}[1]{\protect\rule{.1in}{.1in}}

\begin{document}

\title{Do confinement and darkness have the same conceptual roots?}
\author{Bert Schroer\\CBPF, Rua Dr. Xavier Sigaud 150, 22290-180 Rio de Janeiro, Brazil\\and Institut fuer Theoretische Physik der FU Berlin, Germany}
\date{February 2008}
\maketitle

\begin{abstract}
In this note the connection between "perfect dark matter" and
string-localization as the tightest noncompact localized matter is pursued in
the context of Yang Mills theories. A programmatic approach based on the
observation that behind the serious infrared problems which prevented the use
of the axial gauge there is unique covariant semiinfinite string-like field,
which flucuates in the endpoint as well as in the string direction, is
formulated and the open problems are explained.

Whereas hitherto the gauge formalism in its local gauge invariants accounted
for the observed hadrons the new formalism based on the consistent use of
string-localized vectorpotentials has a good chance to unravel the nonlocal
part in which energy-momentum carrying dark/confined objects (gluons,
quarks,..) may be residing.

\end{abstract}

\subsection{Positive energy matter with string-localized generators}

In a previous paper the question of whether perfect dark matter (pDM) is
consistent with point-like QFT was investigated \cite{DM1}. Here pDM matter is
a hypothetical positive energy matter which coexists together with standard
matter is part of one joint theory but which cannot be created from collisions
of matter as we know it. Hence the observational relevance of the present work
depends on a null-effect in the planned DM creation experiments. As a result
of the many connections to basic structures of QFT, in particular unsolved
problems of gauge theory, some of the theoretical problems which pDM raises
are also of an autonomous theoretical interest.

Our main finding in \cite{DM1} was that it is not possible to emulate pDM if
it would be, like standard matter, generated by point-like fields. The crucial
theorem in this regard states that in higher dimensional QFT ($d\geq4$) it is
not possible, barring the weird case of a doubled universe with photons which
only couple to one tensor factor, to have a creation of exotic matter from
standard matter if the exotic matter is also generated by point-like fields. \ 

This directs the attention to semiinfinite string-like generated matter (SGM).
Before we remind the reader of some old but little known rigorous results on
this matter, it is necessary to emphasize (in case it was not noticed already)
that string theory (ST) is not SGM \cite{crisis}; in fact the quantization of
the classical Nambu-Goto relativistic string leads to a point-like localized
quantum field; the classical string degrees of freedom go into an infinite
mass tower which is \textit{localized over the same point}. With other words
the characteristic feature of this model (as compared to standard QFT which
results from the quantization of classical point-like fields) consists in an
enormous spatial agglomeration of degrees of freedom over one point. Since the
(graded in presence of supersymmetry) commutator is a c-number, the N-G
"string" field is technically speaking a "generalized free field" with an
infinite mass and spin spectrum\footnote{It would be interesting to understand
whether the geometrically introduced interactions (splitting and recombination
of tubes) admit a reasonable representation in terms of introducing
interactions on generalized free fields.}. As a result one finds all the
properties of generalized free fields which quantum field theorists consider
as unphysical as Hagedorn thermal behavior, breakdown of the time slice
property \cite{C-F} and other results of mental games for which there is no
indication in nature.

The excuse (using Feynman's reproach of ST) of string theorists, who have
looked at the issue of localization, is that the string field only reveal the
visible part of the string which only consists the c.o.m. point, i.e. they
attribute an intrinsic meaning to the often classical way in which
quantization is started\footnote{Whereas it is true that a functional
integrals or the canonical Lagrangian quantization are useful for constructing
quantum theories, one has to be very careful to base conclusions about
localization properties on functional integrands. Causal localization has a
totally autonomous meaning on the quantum side \cite{DM1} which accidentally
agrees with the classical localization in the point-like case. Without this
coalescence Jordan would not have had the chance to discover field
quantization before Wigner's 1939 intrinsic access to particles.}. This
metaphoric way falls back behind Heisenberg's notion of observables which was
introduced just for the purpose of avoiding potentially misleading
(quasi)classical metaphors. Every reader who understands the definition of SMG
matter will not fail to realize that string theory does not deal with SGM.
With this reverence to ST, we are now ready to continue our presentation of
SGM without making appologies everytime we use the word "string".

QFT supports the existence of massive SGM, since the standard mass gap
assumption which led to the derivation of (LSZ, Haag-Ruelle) scattering theory
from the principles of QFT also permits to prove that the field sectors
associated to local observables are at worst string-localized \cite{BF}%
\cite{DM1}; i.e. no "branes" or other higher dimensional localizations are
necessary to generate the full content of QFT. Of course one is free to to
consider even in point-like QFT decomposable strings (and higher dimensional
objects) by integrating point-like fields along strings/submanifolds, although
there are no structural reasons for doing this (quantum interpretation of
quasiclassical solutions,...)

But the theoretical research around this Buchholz-Fredenhagen theorem has been
hampered by the apparent lack of perturbative insight; in addition it is very
difficult to find any observable ramifications of these massive algebraic SGM
in the energy-momentum spectrum and localization structure of particle states.
The biggest on-shell change caused by strings is the breakdown of the crossing
property\footnote{The on-shell crossing property requires the existence of
pointlike interpolating fields for the respective particles.} and its
corrolar, the Aks theorem \cite{Aks}, which states that there can be no
elastic scattering without particle creation.

This leaves the hypothetical possibility of pDM which cannot be created from
collisions from ordinary (point-like generated) matter \cite{DM1}. In fact the
authors speculated at the end of their paper \cite{BF} that this could be a
field theoretic mechanism for confinement in the sense of a vanishing creation
cross section for string-localized matter\footnote{In later work \cite{Bpart}
quarks and gluons were identified with "ultraparticles" i.e. objects which do
not exist in the actual theory but rather in a canonically associated scaling
theory.}. It is not far-fetched to assume that the authors at that time did
not think in terms of pDM which seems to provide a stronger motivation for the
absence of production via evasion of the Aks theorem. The apparent
nonperturbative nature of BF strings has been an impediment to their exploration.

This leaves the possibility to start from string-localized zero mass free
fields and study the perturbative interaction of such fields. The Wigner
positive energy representation theory classifies such starting points, namely
the countable family of finite helicity representations and the continuous
family of "infinite spin" representations. Of these two families only the
infinite spin family is \textit{string-localized as a representation} in the
sense that there are no point-like generated subspaces.

The shortcomings of the use of such representations (possible lack of a
point-like energy-stress tensor) for pDM have been presented in a previous
paper \cite{DM1} where it was pointed out that the unusual properties of this
kind of matter require equally unusual properties of the still elusive QG in
order to couple in a mutually compatible way. So the string-localized
generators of the infinite spin free field are nice mathematical illustration
of perfect string localization and darkness, but they do not fit into our
present picture of particle physics.

On the other hand the finite helicity family can be generated by point-like
field strengths\footnote{The electromagnetic field strength for s=1 and the
4$^{th}$ order tensor field strength (with the tensor symmetry of the Riemann
tensor) for s=2 and the related vector- respectively metric tensor-
potential.} but there is a significant difference to the situation of massive
higher spin field. Whereas the latter are described in the spinorial formalism
by undotted/dotted fields of the form \cite{Wei}\cite{DM1}%

\begin{equation}
\Phi^{(A,\dot{B})}(x),\text{ \ }~\left\vert A-\dot{B}\right\vert \leq s\leq
A+\dot{B}\label{sp}%
\end{equation}
where all infinitely many pairs ($A,\dot{B}$) in the given range are realized,
the zero mass situation is only consistent with those (still infinitely many)
special realizations obeying $A=\dot{B}.$ Luckily one can regain the lost
possibilities by allowing string-like localized generators. Hence the
important vectorpotentials for s=1 and the metric tensor potential for s=2
which would violate this rule are described by string fields $A_{\mu}(x,e)$
and $g_{\mu\nu}(x,e)$ which are localized on the semiinfinite string
$x+\mathbb{R}_{+}e~$\cite{Jens}$.$%
\begin{align}
&  \left(  \Omega,A_{\mu}(x;e)A_{\nu}(x^{\prime};e^{\prime})\Omega\right)
=\frac{1}{\left(  2\pi\right)  ^{3/2}}\int d\mu(p)e^{i(x-x^{\prime})p}%
M_{\mu\nu}(p;e,e^{\prime}),~d\mu(p)=measure~on~\partial V_{+}\label{A-string}%
\\
&  M_{\mu\nu}(p;e,e^{\prime})=-g_{\mu\nu}-\frac{p_{\mu}p_{\nu}(e\cdot
e^{\prime})}{(e\cdot p-i\varepsilon)(e^{\prime}\cdot p+i\varepsilon^{\prime}%
)}+\frac{p_{\mu}e_{\nu}}{(e\cdot p-i\varepsilon)}+\frac{p_{\nu}e_{\mu}%
}{(e^{\prime}\cdot p+i\varepsilon^{\prime})}\nonumber
\end{align}%
\begin{align}
\left(  \Omega,g_{\mu\nu}(x,e)g_{\kappa\lambda}(x^{\prime},e^{\prime}%
)\Omega\right)   &  =\frac{1}{\left(  2\pi\right)  ^{3/2}}\int d\mu
(p)e^{i(x-x^{\prime})p}M_{\mu\nu\kappa\lambda}(p;e,e^{\prime})\label{g-string}%
\\
M_{\mu\nu\kappa\lambda}(p;e,e^{\prime}) &  =M_{\mu\sigma\nu\tau\kappa
\alpha\lambda\beta}^{R}(p)\frac{e^{\sigma}e^{\tau}(e^{\prime})^{\alpha
}(e^{\prime})^{\beta}}{(e\cdot p-i\varepsilon)^{2}(e^{\prime}\cdot
p+i\varepsilon^{\prime})^{2}}\nonumber
\end{align}
Here the summation convention has been used and $M^{R}$ stands for the
two-point function of the point-like spin 2 field strength (a 4$^{th}$ degree
homogeneous polynomial in p) which is identical to that of a linearized
Riemann tensor. These string-localized fields are covariant in the sense that
the string direction $e$ transforms under the homogeneous part of the
Poincar\'{e} group. Free string-like potentials (from which one obtains the
point-like field strength by differentiation) can also directly written in
terms of explicitly calculated interwiner u(p,e) \cite{MSY}. The full
$(A,\dot{B})$ spectrum (\ref{sp}) can be recuperated by allowing
string-localization; furthermore string-like localization for both massive and
massless spin s fields are expected to overcome the van Dam-Veltman No-Go
theorem for approaching massless linearized gravity from its massive version
via point-like fields \cite{Velt}.

Different from the gauge theoretic setting, the string-localized potentials
are operator-valued distributions in both $x$ and the spacelike unit vector
$e$ i.e. the field fluctuates in both, so $e$ does not behave as a gauge
parameter but as a localization point in 1+2 dimensional de Sitter space.
Although there is no variation along the string, the situation is not that of
an object "living" on the tensor product of the Minkowki spacetime and the
3-dim. de Sitter space of spacelike unit vectors e. In other words the
commutativity of the strings is not following the rules of a tensor product of
Minkowski- and the de Sitter- space but is rather decided by the spacelike
separation of the two semiinfinite strings $x+\mathbb{R}_{+}e$ and
$x\prime+\mathbb{R}_{+}e^{\prime}$.

An important property which makes a field description fit for perturbative
renormalization is its short distance scale dimensions (sdd). Although the
field strengths have $sdd\geq2$ which increase with helicity and are already
for $s=1$ above the power counting limit, the sdd of the potentials stays at
the smallest possible value namely $sdd=1$. The metaphoric explanation of this
phenomenon is that the linear transition from point-like field strength to
string-like potentials by formal integrations along a semiinfinite line
encodes part of the short distance fluctuations into directional
e-fluctuations i.e. into Minkowski space infrared fluctuations (which are
short distance in the de Sitter sense).

The short distance improving feature of string-localization is of course also
effective for massive fields. In that case even the s=1 vector field has
$sdd=2$ which surpasses the power counting limit; string-localization brings
it back to renormalization friendly value $sdd=1$. Only if one starts from
zero mass the Schwinger-Higgs screening mechanism is implemented: the complex
scalar field with the wrong sign of the mass term is converted into a
point-localized neutral field instead (as for a usual mass term) of a
delocalized (with string-localized generators) charged field surrounded by
photons. But a calculation which starts with free massive vectormesons and
forces them to lower their $sdd=2$ to $sdd=1$ by combining them with massive
BRST ghosts (i.e. using the sdd improving aspect of the BRST setting which is
not restricted to gauge theories) shows that the Schwinger-Higgs screening
picture is a helpful metaphor for the apparent necessity that interacting
point-like local massive vectormesons apparantly must be be accompanied by a
scalar massive companion (naturally without requiring a "vacuum
condensate")\footnote{This vagely resembles particles to come in multiplets in
the supsersymmetric setting of QFT exept that in the present it is the
locality principle and not group theory which groups particles together} thus
reducing the Higgs phenomenon to the well-known locality principle. A
formulation in terms of a string-localized massive vectorpotential without
BRST ghosts (and naturally without Higgs condensates for which there is no
purpose in a perturbative approach which starts already with "fat"
vectormesons) would remove the remaining doubts about whether the Higgs
phenomenon is part of a more intrinsic phenomenon that massive higher spin
objects (in this case s=1) require the company of lower spin companions (in
this case s=0) in order to uphold locality.

Returning to the zero mass vectorpotentials, it is not difficult to see that
the string-localized vectorfield (\ref{A-string}) is formally identical to the
axial gauge as a result of the second of the two relations $\partial^{\mu
}A_{\mu}(x,e)=0=e^{\mu}A_{\mu}(x,e).$ But the construction of string-localized
fields has nothing to do with gauge fixing; it is rather the result of the
existence of a string-localized vectorpotential which is the only covariant
vectorfield in the Wigner-Fock space. A closely related remark is that since
the string-localized potential lives in the physical Wigner-Fock space, there
is no reason for removing unphysical ghosts; at this point it is also helpful
to remind oneself that the gauge setting does not really address symmetries
but rather converts a redundancy resulting from enforcing an apparent
point-like description via the BRST formalism into a (fake from a physical
viewpoint) symmetry so that the perturbative physics can be retrieved via the
associated cohomology.

Finally one should also point out that the axial gauge has only been used in
formal canonical arguments \cite{axial} and never successfully in a
perturbative renormalization approach; it is well-known that any attempt
failed as a result of severe infrared divergencies. The string-localized field
"explains" the origin of this problem in terms of fluctuations of the string
direction and indicates (see the next section) how to formulate a new
renormalization theory which takes care of string fluctuations.

For experts of local quantum physics who are familiar with the content of
Haag's book \cite{Haag} it may be interesting to point out that, whereas the
local algebras $\mathcal{A(O})$ generated by massive free fields are Haag dual
for arbitrary connected causally complete regions $\mathcal{O},$ this does not
hold for the algebras generated by zero mass field strengths where the duality
only holds for simply connected region but is violated for multiply connected
$\mathcal{O}s$ \cite{Ro}\footnote{In Rumsfeldian terminology this is the best
example I know for illustrating what means an "unknown known" in QFT.}. In
such a case the algebra associated to the multiplically connected region has
more operators than those which can be locally generated within that region.
The existence of string-localized potentials in some way "explains" this
phenomenon \cite{MSY}.

\section{Towards a perturbative renormalization for massless string-localized
vector potentials}

The only interacting string-localized QFT which has been reasonably well
understood in terms of a perturbative renormalizable gauge setting is QED. In
the standard treatment where one uses Feynman rules for the gauge dependent
point-like matter and vectorpotential field the string localization is not yet
visible but any attempt to define a physical charged field shows that this is
only possible by excepting delocalization. The best one can do (in the sense
of sharpest localization) is to define semiinfinite string-localized charged
fields of the Jordan-Mandelstam form%
\begin{align}
\Psi(x,\mathbf{e}) &  =\psi(x)expie\int_{x^{\prime}=x+\mathcal{R}%
_{+}\mathbf{e}}A_{\mu}(x^{\prime})dx^{\prime\mu}\label{Jor}\\
\Psi(x,f) &  =\psi(x)expie\int f^{\mu}(x-x^{\prime})A_{\mu}(x^{\prime
})dx^{\prime\mu},~\partial_{\mu}f^{\mu}(x)=\delta^{4}(x)\nonumber
\end{align}

Steinmann has succeeded to show that this formula (in the smearing sense of
the second line) can be given a rigorous perturbative meaning. His arguments
are quite tedious and involve a new organization of perturbation theory; the
reader is referred to his paper \cite{Steinm}.

Less string-like and more Coulombian smearings leads to other less covariant
delocalizations. Common to all physical charge carrying operators is the
infra-particle nature of their energy momentum spectrum which consists of a
mass-shell singularity which (due to the very strong interaction of charged
particles with infrared photons) has been sucked into the photon continuum, a
fact which can be most easily observed in the Kallen-Lehmann representation of
the two-point function of the charged field. The spectrum starts at the alias
mass shell $p^{2}=m^{2}$ with a sub delta function power behavior which is
charateristic for infraparticles (which were recently rediscovered and called
unparticles). These infraparticles are despite their string-like localization
the best "candles" of particle physics, they consist of a hard nucleus which
however cannot be separated from its soft photonic halo which is responsible
for the delocalization and at the same time for the continuous emission of
infrared photons.

One could be satisfied with this gauge theoretic description if it would not
be for the fact that the construction of the physical charge fields has to be
done "by hand". A formula as the above one is not suggested by an intrinsic
canonical procedure of the Lagrangian or causal approach but rather is a
desperate attempt to define and control the most important objects of QED
namely the charged particles and their physical properties.

This ad hoc way becomes a serious problem in the case of nonabelian gauge
theories. Take the simplest case of SU(2) gluons; it is impossible to find any
perturbatively renormalized mathematical controlled proposal for a gauge
invariant nonlocal object, one only knows how to formally construct point-like
gauge invariant gluonium composites and their may be people who convert their
inability to come up with a conjecture into the far-fetched claim that there
are no nonlocal gauge invariants in nonabelian theories. Hence one would like
to have a setting in which such problems can be addressed in a more natural way.

This background is the main point of departure for a new pragmatic approach to
string localization and in particular to the problem of a possible
string-localized matter content of Yang-Mills theories which may carry
unobserved but gravitationally relevant energy-momentum.

The point of departure of a perturbative attempt is this direction are the
string-localized covariant vector potentials obtained from the Wigner
representation theory. In the following (as well as in previous work)
"string-localized" always refers to generating fields $A(x,e)$ which are
localized along a spacelike half-lines $x+\mathbb{R}_{+}e$.

A formulation of perturbation theory in which causal localization plays a
central role is the Stueckelberg-Bogoliubov-Epstein-Glaser setting also
referred to as "causal perturbation theory". Its input is a point-like scalar
composite field (the interaction Lagrangian) $\mathcal{L}_{int}$ which is a
Wick-ordered polynomial in free fields\footnote{By honoring this historical
notation we do not want it to be misred as being part of a Lagrangian
description. In causal perturbation theory, different from the functional
integral setting, there is no need for a Lagrangian formalism and most of the
different ways of relating the unique Wigner theory with a covariant spinorial
description (\ref{sp}) are not of the Euler-Lagrange type. This approach is
intrinsic i.e. not dependent on any Lagrangian paralellism.}. For the case at
hand these free fields may also be string-localized vectorpotentials $A_{\mu
}(x,e)$. As in recent treatments of gauge theories and QFTs in CST it has been
customary to limit the interaction to a compact spacetime region in order to
concentrate on the algebraic structure and avoid infrared problems caused by
the inappropriateness of global states in the perturbative Wigner-Fock space.
The construction is inductive i.e. knowing the time ordered product of $n$
$\mathcal{L}_{int}$ one tries to determine the $(n+1)^{th}$ order including an
explicit parametrization of the possible ambiguities (renormalization). The
most important concept in this step is the causal locality and the notion of a
minimal scaling degree. For point-like fields in Minkowski spacetime the
validity of translational invariance greatly simplifies the determination of
the ambiguity; as a result of its support on the joint diagonal on the n+1
spacetime points it only involves (derivatives of ) delta functions and the
requirement of a \textit{minimal scaling degree} limits the degree of the
delta functions. In the case of CST the lack of translation invariance makes
the discussion more involved; in particular the minimal degree implementation
requires the analysis of a degree associated with submanifolds with the help
of microcausal analysis.

The complication of lack of translation invariance does not occur with
string-localized fields, but the time ordering of strings and the problem of
the minimal scaling degree is more involved. The string-configurations which
violate the causal separations in the sense of Bogoliubov turn out to form an
open set whereas in the point-like case they just consist of a point (the
complete diagonal) problem at hand namely renormalization involving
string-localized fields requires a generalization which is presently under
investigation. Since in that case even in the presence of translational
invariance the problem of time ordering is rather involved (the strings which
are in the causally non-separated position in the sense of the Bogoliubov
approach form an open set instead of a point). There are simpler
subconfigurations which are characterized by a timelike vector for which the
strings lie in a 3-dim hyperplane where it is clear that the mentioned
Bogoliubov region consists of strings which are mutually included. These
restriction of correlations to such subregions make sense if the distributions
in $e$ allow a restriction to a 3 dimensional hypersurface (corresponding to a
2-dimensional de Sitter subspace) which seems to be the case. The general
situation may then be build up by starting from these subconfigurations in
$\mathbb{R}^{n(4\times3)}$ and their scaling degrees. A programmatic sketch is
however no substitute for concrete results.

There is a significant difference between the QED type situation where
vectorpotentials are coupled to (massive) matter fields of $s$ $\leq1/2$ and
the Yang-Mills type mutual interactions between vectormesons and possibly
interactions with low spin matter fields. In the first case the point-like
locality between a free string-like vectorpotential and its linearly related
field strength is preserved in interactions since the directional dependence
from internal vectormeson propagators can be (by partial integrations)
removed, using the fact that the directional change is of the derivative form
and drops out after moving the derivative to the conserved current (pretty
much as in the gauge setting). The correlation functions involving matter
fields on the other hand will have a directional dependence through radiative
corrections and by forcing the directions in the above indicated way into a
small neighborhood around one preferred direction as mentioned before, we
describe a situation similar Jordan-Mandelstam formula (\ref{Jor}).

The Yang-Mills situation is totally different. In that case the requirement to
find a point-like subalgebra for a theory of interacting string-like
vectormesons is very restrictive. Already the requirement that the spacetime
integral over $\mathcal{L}_{int}$ is formally directional independent leads to
a restriction which however does not yet secure the existence of local
subalgebras generated by nonlinear composites. In fact, taking a hint from the
gauge setting, the additive change under a directional change in the free case
should change and the higher order contributions should build up to a
multiplicative rotation in the component space of the vectormesons with a
composite field of 4$^{th}$ degree being the lowest point-like composite.
There is no group theory "by hand" involved here, since the only principle one
has in this setting is the existence of local observables within a setting of
coupled string-like vectorpotentials; i.e. the gauge classical "crutches" have
been lost. Last not least this method should produce non-local gauge
invariants for which the "by hand" method failed.

We expect that the gauge- and the string- setting coalesce on the local
observables \cite{DF}\cite{Hollands}, and we anticipate that the limitation of
the perturbation theory to the construction of the local algebras and its
ineptness for the construction of states is common to both approaches.

The new approach should be first tested in QED because in that case we already
know a lot about string-localized generators from the mathematical
reformulation of the Jordan-Mandelstam conjecture. The advantage of such an
approach is obvious, the construction of electric charge-carrying operators is
intrinsic and not done by hand i.e. it does not depend on the ingenuity of
guessing gauge invariant nonlocal expressions.

\section{String-localization, confinement and darkness}

The fact that electrically charged infraparticles are our best "candles" shows
that it is ill-advised to base a DM description on the energy-momentum
spectrum which is identical to that of the "unparticle" proposal
\cite{Geo}\cite{infra}. Since pDM can only occur in string-localized sectors
and since we (at least temporarily) rejected the only explicitly known model
namely the string-localized energy-momentum carrying infinite spin field on
the ground of its apparent extreme inertness (even with respect to
semiclassical Einstein-Hilbert gravity), our search will be narrowed to
(excluding higher helicity matter) selfinteracting massless vectorpotentials.

What lends additional importance to such a search is the fact that virtually
nothing is known about possible \textit{nonlocal gauge invariant sectors of
Yang-Mills theories}. To believe that the nonlocal gauge-invariant richness
generated by gauge invariant string-localized operators suddenly disappears
and only local gluonium and hadronic matter remains, is not very palatable,
although some particle theorists seem to have accepted this view.

Most particle physicists considers quarks and gluons as practically useful but
conceptually unresolved metaphors. Others refer to the scaling limit of QCD
which they imagine to be a free point-like theory of quarks and a collection
of free gluonic field strengths. At this point usually the reference to
asymptotic freedom enters. But one should keep in mind that the asymptotic
freedom statement is a concistency statement and not a theorem\footnote{This
was the point view of Kurt Symanzik who prepared the ground for the
perturbative asymptotic freedom calculation. Whereas e.g. in the O(n)
Gross-Neveu model asymptotic freedom is a genuine property that model, in QCD
this property refers to the consistency between the unknown confinement, whose
renormalization group parameter space are the physical masses (coming from an
ill-understood mass transmutation), and a perturbative accessible short
distance phase in terms of a not directly observable coupling g.}. Even if one
excepts it as a theorem, it would not solve the problem about the reality
content of QCD which in the present context amounts to the possible existence
of non-local string-generated sectors in Yang-Mills theories. At this point
the searches for pDM and a better understanding of confinement of gluons and
quarks coalesce.

Despite its unsolved infrared aspects, the axial gauge still enjoys popularity
in canonical arguments \cite{axial}. It is our conviction \cite{MSY} that
these difficulties will be overcome after realizing that one is not dealing
with a gauge fixing but rather with a unique covariant string-localized
potential in a physical Hilbert space i.e. when one treats these fields as
fluctuating in a bigger space as explained in the previous section.

Whereas for the determination of the renormalization freedom (thinking about a
causal Epstein-Glaser approach extended to string fields) one needs to leave
the string position (with the number of independent string directions which
increases with the perturbative order), for the physical interpretation one
wants to construct the sharpest localized generators by limiting the
directional vacuum fluctuation via a compactly localized test function around
one chosen direction as in (\ref{Jor}). So the first application of the
string-localized formalism should consist in re-doing a Steinmann like
calculation but without writing down a Jordan-Mandelstam Ansatz for charged
fields by having the charged fields emerge from the renormalization formalism.

Whereas in the QED one has clear expectations about the string-like formalism
being generated by charged particles whose permanent infrared photon-dressing
can be compressed into an arbitrarily thin spacelike cone (from where one may
generate more spread out nonlocal situations), the same cannot be said about
Yang-Mills like models in the new formulation. The fact that the relation
between local observables and string-like building blocks\footnote{Since the
computations have not yet been done, we cannot exclude the case that
string-localized potentials can only be used to construct the perturbation
theory of locally observable composites. The resulting situation would rule
out the use of self-coupled vectormesons for the construction of interacting
string-localized matter and pDM, but it would be interesting in its own right
since it may lead to a theorem about the existence of QFT whose only
connection to Lagrangians is via composite fields.} becomes nonlinear,
forecloses any attempt to visualize the strings as line integrals over
(e-independent) local observables. In this situation the expected larger
Hilbert space can not be arrived at by applying local observables; the model
will have additional physical delocalized "stuff" generated by acting with the
interacting vectorpotentials onto the vacuum.

It is this faith that behind the barriers of the infrared divergent axial
gauge setting of Yang-Mills models there is a new string-based formalism
\cite{MS}, which supports our conjecture that confinement and darkness may be
connected. Surely if there are nonlocal generators (in the gauge setting:
nonlocal gauge invariant composites which cannot be obtained from local gauge
invariants), i.e. if the phenomenon of string-like generation is not limited
to abelian theories, they must carry nontrivial energy-momentum, and since the
observed QCD spectrum covers all the seen particles, consistency demands they
must be string-generated and dark.

Some messages can be abstracted from existing perturbative calculations based
on the BRST gauge setting \cite{DF}\cite{Hollands}. These are perturbative
approaches which separate the problem of the construction of the local
algebras from that of states. About the second which includes the particle
content and scattering theory, one can presently say nothing apart from the
statement that these problems are outside of perturbation theory (at least for
YM models). However, as shown in the two papers, the local algebraic structure
can be perturbatively accessed. This turns the problem of finding the physical
states into the mathematical problem of constructing certain states on
operator algebras. The string-localized setting is expected to come with the
same problems apart from the fact that it is expected to do much better on
producing nonlocal (string-localized) interacting operators.

\bigskip

.

\end{document}